\documentclass[aps,twocolumn]{revtex4}
\usepackage{epsf}
\usepackage{graphicx}

\begin{document}

\title
{Quantum-Classical Dynamics of Wave Fields}

\author{Alessandro Sergi
\footnote{E-mail: asergi@unime.it}}
\affiliation{
Dipartimento di Fisica, 
Universit\'a degli Studi di Messina,
Contrada Papardo 98166 Messina, Italy
}

\begin{abstract}
An approach to the quantum-classical mechanics of 
phase space dependent operators, which has been proposed recently, 
is remodeled as a formalism for wave fields.
Such wave fields obey a system of coupled non-linear equations
that can be written by means of a suitable non-Hamiltonian bracket.
As an example, the theory is applied to the relaxation dynamics
of the spin-boson model.
In the adiabatic limit, a good agreement with calculations
performed by the operator approach is obtained.
Moreover, the theory proposed in this paper
can  take nonadiabatic effects into account
without resorting to surface-hopping approximations.
Hence, the results obtained follow qualitatively those
of previous surface-hopping calculations
and increase by a factor of (at least) two the time length 
over which nonadiabatic dynamics can be propagated
with small statistical errors.
Moreover,
it is worth to note that the  dynamics of 
quantum-classical wave fields here proposed is a straightforward
non-Hamiltonian generalization of the formalism for non-linear
quantum mechanics that Weinberg introduced recently.
\end{abstract}

\maketitle


\section{Introduction}

There are many instances where a quantum-classical description
can be a useful approximation to full quantum dynamics.
Typically, a quantum-classical picture often 
allows one to implement calculable algorithms on computers
whenever charge transfer is considered
within complex environments, such as those provided by
proteins or nano-systems in general~\cite{ksreview}.
With respect to this, an algebraic approach has been recently
proposed~\cite{qc-bracket,kcmqc} in order to formulate the dynamics and the
statistical mechanics~\cite{qc-stat} of quantum-classical systems.
General questions regarding the quantum-classical correspondence
have also been addressed within a similar framework~\cite{brumer}.
The approach of Refs.~\cite{qc-bracket,kcmqc}
represents quantum-classical dynamics by means
of suitable brackets of phase space dependent operators
and describes consistently the back-reaction
between quantum and classical degrees of freedom.
Notably, a particular implementation of this
formalism has been used to calculate nonadiabatic rate constants
in systems modeling chemical reactions in the condensed phase~\cite{kapral}.
However, such schemes have only permitted the simulation of
short-time nonadiabatic dynamics
because of the time-growing statistical error of the algorithm.
Nevertheless, the algebraic approach~\cite{qc-bracket,kcmqc},
underlying the algorithms of Refs.~\cite{kapral}, 
has some very nice features, such as the (above mentioned) proper description
of the back-reaction between degrees of freedom, that 
one should not give up
when addressing quantum-classical statistical mechanics. 
Moreover, quantum-classical brackets
define a non-Hamiltonian algebra~\cite{b3} so that
their matrix structure allows one to introduce quantum-classical
Nos\'e-Hoover dynamics~\cite{b3} and to define the
statistical mechanics of quantum-classical systems
with holonomic constraints~\cite{bsilurante}.
All of the above features of the formalism
are highly desirable when studying complex systems
in condensed phases.
Therefore, it is worth to search for a reformulation
of the theory of Refs.~\cite{kapral,b3,bsilurante}
that, while maintaining such features, could be used
to integrate reliably long-time nonadiabatic dynamics. 

To this end, one can note that, within standard quantum mechanics,
some problems that are formidable to solve by means of the dynamics
of operators become much simpler to handle
when, instead, the time evolution of wave functions is considered~\cite{ballentine}.
Hence, for analogy, it might also happen that,
within quantum-classical mechanics,
the correspondence between operators and quantum-classical wave functions
could open new possibilities for useful approximations
in order to carry long-time calculations efficiently.
Indeed, finding and applying the correspondence
between operator and wave scheme of motion in quantum-classical
mechanics is the scope of the present paper.
A wave picture for quantum-classical
dynamics can be found by direct algebraic manipulation of the
equation of motion for the density matrix.
In practice, the single equation obeyed by the quantum-classical density
matrix is mapped onto two coupled non-linear equations
for quantum-classical wave fields.
Despite its non-linear character, such a quantum-classical dynamics of 
phase space dependent
wave fields corresponds exactly to the dynamics
of phase space dependent operators discussed 
in Refs.~\cite{qc-bracket,kcmqc,kapral,b3,bsilurante}
and can be used to devise novel algorithms and approximation schemes.

The abstract algebraic equations here presented are readily expressed
in the adiabatic basis and applied, in order to provide 
an illustrative example, to the spin-boson model
and its relaxation dynamics both in the adiabatic and nonadiabatic limit.
By making a suitable equilibrium approximation to the non-linear wave equations,
it is found that nonadiabatic dynamics can be propagated, within the wave picture,
for time intervals that are a factor of two-three longer than those
which have been spanned  in Ref.~\cite{qc-sb} 
by means of the operator theory~\cite{qc-bracket,kcmqc,kapral,b3,bsilurante}.
Such a result is very encouraging for pursuing the long-time integration
of the nonadiabatic dynamics of complex systems in condensed phases.

Following a line of research that investigates the relations
between classical and quantum theories~\cite{qgen},
it is worth to note that the wave picture of quantum-classical mechanics,
which is introduced in this paper, generalizes within a non-Hamiltonian framework
the elegant formalism that Weinberg~\cite{weinberg}
proposed for describing possible non-linear effects 
in quantum mechanics~\cite{nonlinear}.

This paper is organized as follows. 
In Section~\ref{sec:bracket} the non-Hamiltonian algebra
of phase space dependent operators is briefly summarized.
In Section~\ref{sec:qcwd} the quantum-classical dynamics of operators
is transformed into a theory
for phase space dependent wave fields evolving in time. 
Such a theory for wave fields is also
expressed by means of suitable non-Hamiltonian brackets:
in this way a link is found with the generalization
of Weinberg's non-linear formalism given in Appendix~\ref{app:weinberg}.
More specifically, in Appendix~\ref{app:weinberg},
Weinberg's formalism is briefly reviewed and its 
symplectic structure is unveiled.  Then, this structure
is generalized by means of non-Hamiltonian brackets.
Therefore, one can appreciate how the generalized Weinberg's
formalism establishes a more comprehensive mathematical framework
for non-linear equations of motion, comprising
phase space dependent wave fields as a special case.
In Section~\ref{sec:qcwdab} the abstract non-linear
equations of motion for quantum-classical fields are 
represented in the adiabatic basis and some considerations,
which pertain to the numerical implementation, are made.
By making an equilibrium \emph{ansatz}, in Section~\ref{sec:sb} the non-linear
equations of motion are put into a linear form and the theory is
applied to the spin-boson model.
Section~\ref{sec:conclusions} is devoted to conclusions and perspectives.

\section{Non-Hamiltonian Mechanics of
 Quantum-Classical Operators}\label{sec:bracket}

A quantum-classical system is composed of
both quantum $\hat{\chi}$ and classical $X$ degrees
of freedom, where $X=(R,P)$ is the phase space point,
with $R$ and $P$ coordinates and momenta, respectively.
Within the operator formalism of Refs.~\cite{qc-bracket,kcmqc,b3,bsilurante},
the quantum variables depend from the classical point, $X$,
of phase space.
The energy of the system is defined in terms of
a Hamiltonian operator $\hat{H}=\hat{H}(X)$, which
couples quantum and classical variables, by
$E={\rm Tr}'\int dX \hat{H}(X)$.
The dynamical evolution of a quantum-classical operator $\hat{\chi}(X)$
is given by~\cite{qc-bracket,kcmqc} 
\begin{eqnarray}
\frac{d}{dt} \hat{\chi}(X,t)&=&
\frac{i}{\hbar}
\left[\hat{H},\hat{\chi}(X,t)\right]_{\mbox{\tiny\boldmath$\cal B$}}
-\frac{1}{2}\left\{\hat{H},\hat{\chi}(X,t)\right\}_{\mbox{\tiny\boldmath$\cal B$}}
\nonumber\\
&+&\frac{1}{2}\left\{\hat{\chi}(X,t),\hat{H}\right\}_{\mbox{\tiny\boldmath$\cal B$}}
=\left(\hat{H},\hat{\chi}(X,t)\right)\;,
\label{eq:qcbracket}
\end{eqnarray}
where 
\begin{eqnarray}
\left[\hat{H} , \hat{\chi}\right]_{\mbox{\tiny\boldmath$\cal B$}}
&=&
\left[\begin{array}{cc} \hat{H} & \hat{\chi}\end{array}\right]
\cdot\mbox{\boldmath$\cal B$}\cdot
\left[\begin{array}{c} \hat{H} \\
 \hat{\chi} \end{array} \right]
\label{eq:qlm}
\end{eqnarray}
is the commutator and
\begin{eqnarray}
\{\hat{H},\hat{\chi}\}_{\mbox{\tiny\boldmath$\cal B$}}
&=&
\sum_{i,j=1}^{2N}
\frac{\partial \hat{H}}{\partial X_i}{\cal B}_{i j}
\frac{\partial \hat{\chi}}{\partial X_j}
\label{Lambda}
\end{eqnarray}
is the  Poisson bracket~\cite{goldstein}.
Both the commutator and the Poisson bracket are defined in terms of the antisymmetric matrix
\begin{equation}
\mbox{\boldmath$\cal B$}=\left[\begin{array}{cc}0 & 1\\ -1 & 0\end{array}\right]\;.
\label{B}
\end{equation}
The last equality in Eq.~(\ref{eq:qcbracket})
defines the quantum-classical bracket.
Following Refs.~\cite{b3,bsilurante,sergi},
the quantum-classical law of motion
can be easily casted in matrix
form as
\begin{eqnarray}
\frac{d}{dt} \hat{\chi}&=&\frac{i}{\hbar}
\left[\begin{array}{cc} \hat{H} & \hat{\chi} \end{array}\right]
\cdot\mbox{\boldmath$\cal D$}\cdot
\left[\begin{array}{c} \hat{H} \\ \hat{\chi} \end{array}\right]
\nonumber\\
&=&\frac{i}{\hbar}[\hat{H},\hat{\chi}]_{\mbox{\tiny\boldmath$\cal D$}}\;,
\label{qclm}
\end{eqnarray}
where
\begin{equation}
\mbox{\boldmath$\cal D$}=\left[\begin{array}{cc} 0& 
1-\frac{\hbar}{2i}
\left\{\ldots,\ldots\right\}_{\mbox{\tiny\boldmath$\cal B$}}
\\
-1+\frac{\hbar}{2i}
\left\{\ldots,\ldots\right\}_{\mbox{\tiny\boldmath$\cal B$}}
& 0\end{array}\right]\;.
\label{D}
\end{equation}
The structure of Eq.~(\ref{qclm})
is that of a non-Hamiltonian commutator, which will be defined
below in  Eq.~(\ref{eq:gen-qlm}), and as such
generalizes the standard quantum law of motion~\cite{b3}.
The antisymmetric super-operator $\mbox{\boldmath$\cal D$}$ in Eq.~(\ref{D}) 
introduces a novel mathematical
structure that characterizes the time evolution of quantum-classical
systems.
The Jacobi relation in quantum-classical dynamics is
\begin{equation}
{\cal J}=\left[\hat{\chi},
\left[\hat{\xi},\hat{\eta}\right]_{\mbox{\tiny\boldmath$\cal D$}} 
\right]_{\mbox{\tiny\boldmath$\cal D$}}
+\left[\hat{\eta},\left[\hat{\chi},\hat{\xi}
\right]_{\mbox{\tiny\boldmath$\cal D$}}
\right]_{\mbox{\tiny\boldmath$\cal D$}}
+\left[\hat{\xi},\left[\hat{\eta},\hat{\chi}
\right]_{\mbox{\tiny\boldmath$\cal D$}}
\right]_{\mbox{\tiny\boldmath$\cal D$}}.
\label{qc-jacobi}
\end{equation}
The explicit expression of $\cal J$ 
has been given in Ref.~\cite{b3} where it was shown 
that it may be different from zero at least in some point $X$ of phase space:
for this reason the quantum-classical theory of Refs.~\cite{qc-bracket,kcmqc,b3,bsilurante}
can be classified as a non-Hamiltonian theory.

It is worth to note that the quantum-classical law of motion
in Eq.~(\ref{qclm})
is a particular example of a more general form of quantum mechanics
where time evolution is defined by means of non-Hamiltonian commutators.
The non-Hamiltonian commutator between two arbitrary operators
$\hat{\chi}$ and $\hat{\xi}$ is defined by
\begin{equation}
[\hat{\chi},\hat{\xi}]_{\mbox{\tiny\boldmath$\Omega$}}=
\left[\begin{array}{cc}\hat{\chi}
& \hat{\xi}\end{array}\right]
\cdot\mbox{\boldmath$\Omega$}\cdot
\left[\begin{array}{c}\hat{\chi} \\
\hat{\xi}\end{array}\right]\;,
\label{eq:gen-quantum-algebra}
\end{equation}
where $\mbox{\boldmath$\Omega$}$ is an antisymmetric matrix operator
of the form
\begin{equation}
\mbox{\boldmath$\Omega$}=
\left[\begin{array}{cc}0 & f[\hat{\eta}]\\ -f[\hat{\eta}] & 
0\end{array}\right]\;,
\end{equation}
where $f[\hat{\eta}]$ can be another arbitrary operator or functional
of operators.
Then, generalized equations of motion can be defined as
\begin{eqnarray}
 \frac{d\hat{\chi}}{dt}&=&\frac{i}{\hbar}
\left[\begin{array}{cc} \hat{H} & \hat{\chi} \end{array}\right]
\cdot\mbox{\boldmath$\Omega$}\cdot
\left[\begin{array}{c} \hat{H} \\ \hat{\chi}\end{array}\right]
\nonumber\\
&=&\frac{i}{\hbar}
[\hat{H} , \hat{\chi}]_{\mbox{\tiny\boldmath$\Omega$}}
\;.
\label{eq:gen-qlm}
\end{eqnarray}
The non-Hamiltonian commutator of Eq.~(\ref{eq:gen-quantum-algebra})
defines a generalized form of quantum mechanics where, nevertheless,
the Hamiltonian operator $\hat{H}$
is still a constant of motion because of the antisymmetry of 
$\mbox{\boldmath$\Omega$}$.

\section{Quantum-classical wave dynamics}\label{sec:qcwd}

In Refs.~\cite{qc-bracket,kcmqc}, quantum-classical evolution
has been formulated in terms of phase space dependent operators.
In this scheme of motion operators evolve according to
\begin{eqnarray}
\hat{\chi}(X,t)&=&\exp\left\{t
\left[\hat{H},\ldots\right]_{\mbox{\tiny\boldmath$\cal D$}}\right\}
\hat{\chi}(X)\nonumber\\
&=&\exp\left\{it{\mathcal L}\right\}\hat{\chi}(X)\;,
\label{eq:qc-heisenberg}
\end{eqnarray}
where the last equality defines the quantum-classical Liouville
propagator.
Quantum-classical averages are calculated as
\begin{eqnarray}
\langle\hat{\chi}\rangle(t)
&=&{\rm Tr}'\int dX\hat{\rho}(X)\hat{\chi}(X,t)
\nonumber\\
&=&{\rm Tr}'\int dX\hat{\rho}(X,t)\hat{\chi}(X)\;,
\label{eq:qc-average}
\end{eqnarray}
where $\hat{\rho}(X)$ is the quantum-classical density matrix
and $\hat{\rho}(X,t)=\exp\left\{-it{\mathcal L}\right\}\hat{\rho}(X)$.
Either evolving the dynamical variables or the density matrix, one is
still dealing with phase space dependent operators:
\emph{viz.}, one deals with a form of generalized
quantum-classical matrix mechanics.
As it has been discussed in the Introduction, 
this theory has interesting formal features and
a certain number of numerical schemes
have been proposed to integrate the dynamics
and calculate correlation functions~\cite{kapral,qc-sb,num-qc}.
However, these algorithms have been applied with success only to
short-time dynamics because of statistical uncertainties that
grow with time beyond numerical tolerance.
With this in mind, it is interesting to see
which features are found when the quantum-classical theory 
of Refs.~\cite{qc-bracket,kcmqc} is mapped
onto a scheme of motion where phase space dependent wave fields, instead
of operators, are used to represent the dynamics.

As it is well known~\cite{ballentine},
in standard quantum mechanics, the correspondence between
dynamics in the Heisenberg and in the Schr\"odinger picture
rests ultimately on the following operator identity:
\begin{equation}
e^{\hat{Y}}\hat{X}e^{-\hat{Y}}=e^{[\hat{Y},\ldots]}\hat{X}\;,
\label{eq:expid}
\end{equation}
where $[\hat{Y},\ldots]\hat{X}\equiv [\hat{Y},\hat{X}]$.
Thus, in quantum-classical theory, one would like to
derive an operator identity
analogous to that in Eq.~(\ref{eq:expid}).
However, as already shown in Ref.~\cite{qc-stat},
because of the non associativity of the quantum-classical
bracket in Eq.~(\ref{qclm}),
the identity that can be derived is
\begin{eqnarray}
e^{\frac{it}{\hbar}~\left[\hat{H},
\ldots\right]_{\mbox{\tiny\boldmath$\cal D$}}} 
\hat{\chi}&=&
{\cal S}\left(
e^{\frac{it}{\hbar}\overrightarrow{\mathcal H}}\hat{\chi}
e^{-\frac{it}{\hbar}\overleftarrow{\mathcal H}}\right)\;,
\label{eq:qc-ope-ide}
\end{eqnarray}
where the two operators
\begin{eqnarray}
\overrightarrow{\mathcal H}&=&\hat{H}
-\frac{\hbar}{2i}
\left\{\hat{H},\ldots\right\}_{\mbox{\tiny\boldmath$\cal B$}}
\label{eq:hright}\\
\overleftarrow{\mathcal H}&=&\hat{H}
-\frac{\hbar}{2i}
\left\{\ldots,\hat{H}\right\}_{\mbox{\tiny\boldmath$\cal B$}}
\label{eq:hleft}
\end{eqnarray}
have been introduced
and $\cal S$ is an ordering operator which is chosen so that
the left and the right hand side of Eq.~(\ref{eq:qc-ope-ide})
coincide by construction~\cite{qc-stat}, when the exponential operators
are substituted with their series expansion.
The existence of such an ordering problem, and of the ordering operator
$\cal S$, in Eq.~(\ref{eq:qc-ope-ide}) is caused by the Poisson bracket
parts of the operators in Eqs.~(\ref{eq:hright}) and~(\ref{eq:hleft}).
Hence, one can imagine that 
the solution to this problem can be found by dealing properly 
with these parts of the brackets.
To this end, one can
consider the quantum-classical equation of motion for the density matrix
\begin{eqnarray}
\frac{\partial\hat{\rho}}{\partial t}&=&
-\frac{i}{\hbar}
\left[\begin{array}{cc}\hat{H} & \hat{\rho}\end{array}\right]
\nonumber\\
&\cdot&
\left[\begin{array}{cc}
0 & 1-\frac{\hbar}{2i}
\left\{\ldots,\ldots\right\}_{\mbox{\tiny\boldmath$\cal B$}}
\\
-1+\frac{\hbar}{2i}
\left\{\ldots,\ldots\right\}_{\mbox{\tiny\boldmath$\cal B$}}
& 0\end{array}\right]
\cdot
\left[\begin{array}{c}\hat{H}\\\hat{\rho}\end{array}\right]\;.
\nonumber
\\
\label{eq:rhoW}
\end{eqnarray}
As above discussed, in Eq.~(\ref{eq:rhoW}) the ordering problem arises
from the  terms in the right hand side containing
the Poisson bracket operator
 $\left\{\ldots,\ldots\right\}_{\mbox{\tiny\boldmath$\cal B$}}$.
Then, considering the identity
$1=\hat{\rho}\cdot\hat{\rho}^{-1}=\hat{\rho}^{-1}\cdot\hat{\rho}$,
Eq.~(\ref{eq:rhoW}) can be rewritten as
\begin{eqnarray}
\partial_t \hat{\rho}
&=&
-\frac{i}{\hbar}
\left[\begin{array}{cc}\hat{H} & \hat{1}\end{array}\right]
\nonumber\\
&\cdot&
\left[\begin{array}{cc}
0 & 1-\frac{\hbar}{2i}
\left\{\ldots,\hat{\rho}\right\}_{\mbox{\tiny\boldmath$\cal B$}}
\\
-1+\frac{\hbar}{2i}
\left\{\hat{\rho},\ldots\right\}_{\mbox{\tiny\boldmath$\cal B$}}
& 0\end{array}\right]
\cdot
\left[\begin{array}{c}\hat{H}\\\hat{1}\end{array}\right]
\nonumber
\\
&=&
-\frac{i}{\hbar}
\left[\begin{array}{cc}\hat{H} & \hat{\rho}\hat{\rho}^{-1}\end{array}\right]
\nonumber\\
&\cdot&
\left[\begin{array}{cc}
0 & 1-\frac{\hbar}{2i}
\left\{\ldots,\hat{\rho}\right\}_{\mbox{\tiny\boldmath$\cal B$}}
\\
-1+\frac{\hbar}{2i}
\left\{\hat{\rho},\ldots\right\}_{\mbox{\tiny\boldmath$\cal B$}}
& 0\end{array}\right]
\cdot
\left[\begin{array}{c}\hat{H}\\\hat{\rho}^{-1}\hat{\rho}\end{array}\right]
\nonumber
\\
&=&
-\frac{i}{\hbar}
\left[\begin{array}{cc}\hat{H} & \hat{\rho}\end{array}\right]
\cdot
\mbox{\boldmath$\cal D$}_{\mbox{\tiny\boldmath$\cal B$},[\hat{\rho}]}
\cdot
\left[\begin{array}{c}\hat{H}\\
\hat{\rho}\end{array}\right]\;,
\label{eq:rho2}
\end{eqnarray}
where
\begin{eqnarray}
\begin{array}{l}
\mbox{\boldmath$\cal D$}_{\mbox{\tiny\boldmath$\cal B$},[\hat{\rho}]}
=\\
\left[\begin{array}{cc}
0 & 1-\frac{\hbar}{2i}
\left\{\ldots,\ln(\hat{\rho})\right\}_{\mbox{\tiny\boldmath$\cal B$}}
\\
-1+\frac{\hbar}{2i}
\left\{\ln(\hat{\rho}),\ldots\right\}_{\mbox{\tiny\boldmath$\cal B$}}
& 0\end{array}\right]
\end{array}\nonumber\\
\label{eq:Drho}
\end{eqnarray}
The operator $\mbox{\boldmath$\cal D$}_{\mbox{\tiny\boldmath$\cal B$},[\hat{\rho}]}$
in Eq.~(\ref{eq:Drho})
depends from the quantum-classical density matrix, $\hat{\rho}$, itself.
However, if one momentarily disregards this non-linear dependence,
Eq.~(\ref{eq:rho2}) can be manipulated algebraically in order to develop
a wave picture of quantum-classical mechanics.
To this end, one can introduce quantum-classical wave fields, $|\psi(X)\rangle$ and
$\langle\psi(X)|$, and make the following {\it ansatz}
for the density matrix
\begin{eqnarray}
\hat{\rho}(X)&=&\sum_{\iota}w_{\iota}|\psi^{\iota}(X)\rangle\langle\psi^{\iota}(X)|
\;,\label{eq:rho-ansatz}
\end{eqnarray}
where one has assumed that, because of thermal disorder,
there can be many microscopic states $|\psi^{\iota}(X)\rangle$~$({\iota}=1,\ldots,l)$
which correspond to the same value
of the macroscopic relevant observables~\cite{balescu}.
In terms of the quantum-classical wave fields, 
$|\psi^{\iota}(X)\rangle$ and $\langle\psi^{\iota}(X)|$,
and considering the single state labeled by $\iota$,
Eq.~(\ref{eq:rho2}) becomes
\begin{eqnarray}
|\dot{\psi}^{\iota}(X)\rangle\langle\psi^{\iota}(X)|
&+&|\psi^{\iota}(X)\rangle\langle\dot{\psi}^{\iota}(X)|
=\nonumber\\
&-&\frac{i}{\hbar}\left(\hat{H}|\psi^{\iota}(X)\rangle\langle\psi^{\iota}(X)|
\right.\nonumber\\
&+&
\left.|\psi^{\iota}(X)\rangle\langle\psi^{\iota}(X)|\hat{H}\right)
\nonumber\\
&+&\frac{1}{2}
\left(
\left\{\hat{H},\ln(\hat{\rho})\right\}_{\mbox{\tiny\boldmath$\cal B$}}
|\psi^{\iota}(X)\rangle\langle\psi^{\iota}(X)|\right.
\nonumber\\
&-&\left.
|\psi^{\iota}(X)\rangle\langle\psi^{\iota}(X)|
\left\{\ln(\hat{\rho}),\hat{H}\right\}_{\mbox{\tiny\boldmath$\cal B$}}
\right)\;.\nonumber\\
\label{eq:wavematrix}
\end{eqnarray}
Equation~(\ref{eq:wavematrix}) can be written as a system
of two coupled equations for the wave fields~\cite{fckrk}:
\begin{eqnarray}
i\hbar\frac{d}{dt}\vert\psi^{\iota}_{(X,t)}\rangle & =&
\left(\hat{H}-\frac{\hbar}{2i}
\left\{\hat{H},\ln(\hat{\rho}_{(X,t)})\right\}_{\mbox{\tiny\boldmath$\cal B$}}
\right)\vert\psi^{\iota}_{(X,t)}\rangle\nonumber\\
-i\hbar\langle\psi^{\iota}_{(X,t)}\vert\overleftarrow{\frac{d}{dt}}
&=&
\langle\psi^{\iota}_{(X,t)}\vert\left(\hat{H}
-\frac{\hbar}{2i}
\left\{
\ln(\hat{\rho}_{(X,t)}),\hat{H}\right\}_{\mbox{\tiny\boldmath$\cal B$}}
\right)
\;.\nonumber\\
\label{eq:fckrk}
\end{eqnarray}
Equations~(\ref{eq:fckrk}), which are obeyed by the wave fields,
are non-linear since their solution depends self-consistently
from the density matrix defined in Eq.~(\ref{eq:rho-ansatz}).
These equations are also non-Hermitian since the operators
$\left\{\hat{H},\ln(\hat{\rho})\right\}_{\mbox{\tiny\boldmath$\cal B$}}$
and $\left\{\ln(\hat{\rho}),\hat{H}\right\}_{\mbox{\tiny\boldmath$\cal B$}}$
are not Hermitian.
However, this does not cause problems for the conservation of probability.
The wave fields
$\vert\psi^{\iota}\rangle$ and $\langle\psi^{\iota}\vert$
evolve according to the different propagators
\begin{eqnarray}
\overrightarrow{\cal U}_{{\mbox{\tiny\boldmath$\cal B$}},[\hat{\rho}]}(t)
&=&
\exp\left[-\frac{it}{\hbar}\left(\hat{H}
-\frac{\hbar}{2i}
\left\{\hat{H},\ln(\hat{\rho})\right\}_{\mbox{\tiny\boldmath$\cal B$}}
\right)\right]\;,\nonumber\\
&&\\
\overleftarrow{\cal U}_{{\mbox{\tiny\boldmath$\cal B$}},[\hat{\rho}]}(t)
&=&
\exp\left[-\frac{it}{\hbar}
\left(\hat{H}
-\frac{\hbar}{2i}
\left\{\ln(\hat{\rho}),\hat{H}\right\}_{\mbox{\tiny\boldmath$\cal B$}}
\right)\right]\;,\nonumber\\
\end{eqnarray}
so that time-propagating wave fields are defined by
\begin{eqnarray}
\vert\psi^{\iota}(X,t)\rangle&=&
\overrightarrow{\cal U}_{{\mbox{\tiny\boldmath$\cal B$}},[\hat{\rho}]}(t)
\vert\psi^{\iota}(X)\rangle
\\
\langle\psi^{\iota}(X,t)\vert&=&
\langle\psi^{\iota}(X,)\vert
\overleftarrow{\cal U}_{{\mbox{\tiny\boldmath$\cal B$}},[\hat{\rho}]}(t)
\;.
\end{eqnarray}
Quantum classical averages can be written as
\begin{eqnarray}   
\langle\hat{\chi}\rangle(t)&=&\int dX\sum_{\iota}w_{\iota}
\langle\psi^{\iota}(X,t)| \hat{\chi} |\psi^{\iota}(X,t)\rangle \;.
\label{eq:wave-ave}
\end{eqnarray}   
One can always transform back to the operator picture
to show that the probability is conserved.


\subsection{Non-linear wave dynamics by means of non-Hamiltonian brackets}

The wave equations in~(\ref{eq:fckrk}) were derived starting
from the non-Hamiltonian commutator expressing
the dynamics of phase space dependent operators~\cite{b3}.
It is interesting to recast quantum-classical wave dynamics itself
by means of non-Hamiltonian brackets.
It turns out that this form of the wave equations
generalizes the mathematical formalism
first proposed by Weinberg~\cite{weinberg} in order
to study possible non-linear effects in quantum mechanics
(see Appendix~\ref{app:weinberg}).

Consider a case in which a single state is present, \emph{i.e.}
$\iota=1$. Then, consider the wave fields $\vert\psi\rangle$
and $\langle\psi\vert$ as coordinates of an abstract space,
and denote the point of such a space as
\begin{equation}
\mbox{\boldmath$\zeta$}=\left[\begin{array}{c}|\psi\rangle\\
\langle\psi|\end{array}\right]\;.
\end{equation}
Introduce the function
\begin{equation}
{\cal H}=\langle\psi\vert\hat{H}\vert\psi\rangle\;,
\end{equation}
and the antisymmetric matrix operator
\begin{equation}
\mbox{\boldmath$\Omega$}
=
\left[\begin{array}{cc}0 &
1-\frac{\hbar}{2i}\frac{\left\{\hat{H},\ln(\hat{\rho})\right\}_{
\mbox{\tiny\boldmath$\cal B$}}\vert\psi\rangle}{\hat{H}\vert\psi\rangle}
\\
-1+\frac{\hbar}{2i}\frac{\left\{\ln(\hat{\rho}),\hat{H}\right\}_{
\mbox{\tiny\boldmath$\cal B$}}\vert\psi\rangle}{\langle\psi\vert\hat{H}}
& 0
\end{array}\right]
\end{equation}
Equations~(\ref{eq:fckrk}) can be written in compact form as
\begin{eqnarray}
\frac{\partial\mbox{\boldmath$\zeta$}}{\partial t}
&=&-\frac{i}{\hbar}
\left[
\begin{array}{cc}\frac{\partial{\cal H}}{\partial\vert\psi\rangle}
&\frac{\partial{\cal H}}{\partial\langle\psi\vert}
\end{array}
\right]
\cdot\mbox{\boldmath$\Omega$}\cdot
\left[\begin{array}{c}\frac{\partial\mbox{\boldmath$\zeta$}}{\partial\vert\psi\rangle}
\\\frac{\partial\mbox{\boldmath$\zeta$}}{\partial\langle\psi\vert}\end{array}\right]
\nonumber\\
&=&-\frac{i}{\hbar}
\left\{{\cal H},\mbox{\boldmath$\zeta$}\right\}_{\mbox{\tiny\boldmath$\Omega$};\zeta}\;.
\label{eq:wein-like}
\end{eqnarray}
Equations~(\ref{eq:fckrk}),
or their compact ``Weinberg-like'' form in Eq.~(\ref{eq:wein-like}),
express the wave picture for the quantum-classical dynamics
of phase space dependent quantum degrees of freedom~\cite{qc-bracket,kcmqc}.
Such a wave picture makes one recognize the intrinsic non-linearity
of quantum-classical dynamics.
This specific features will be discussed, among other issues, in the next
section.

\section{Adiabatic basis representation and surface-hopping schemes}\label{sec:qcwdab}

Equations~(\ref{eq:fckrk}) are written in an abstract form.
In order to devise a numerical algorithm to solve them,
one has to obtain a representation in some basis.
Of course, any basis can be used but, since one would like to
find a comparison with surface-hopping schemes, the adiabatic 
basis is a good choice.
To this end, consider the following form of the quantum-classical
Hamiltonian operator:
\begin{equation}
\hat{H}=\frac{P^2}{2M}+\hat{h}(R)\;,
\end{equation}
where the first term provides the kinetic energy of the classical
degrees of freedom with mass $M$, while $\hat{h}(R)$ describes
the quantum sub-system and its coupling with the classical
coordinates $R$.
The adiabatic basis is then defined by the following eigenvalue
equation:
\begin{equation}
\hat{h}\vert\alpha;R\rangle=E_{\alpha}(R)\vert\alpha;R\rangle\;.
\end{equation}
Since the non-linear wave equations in~(\ref{eq:fckrk})
have been derived from the bracket equation for the
quantum-classical density matrix~(\ref{eq:rhoW}),
by dealing in a suitable manner with the
Poisson bracket terms, the most simple way
to find the representation of  the wave equations~(\ref{eq:fckrk})
in the adiabatic basis
is to first represent Eq.~(\ref{eq:rhoW}) in such a basis
and then deal with the terms arising from the Poisson brackets.
The adiabatic representation of Eq.~(\ref{eq:rhoW}) is~\cite{kcmqc}
\begin{eqnarray}
\partial_t \rho_{\alpha\alpha^{\prime}}(X,t)
&=&-\sum_{\beta\beta^{\prime}}
i{\cal L}_{\alpha\alpha^{\prime},\beta\beta^{\prime}}
\rho_{\beta\beta^{\prime}}(X,t) \;,
\end{eqnarray}
where
\begin{eqnarray}
i{\cal L}_{\alpha\alpha^{\prime},\beta\beta^{\prime}}
&=&
i{\cal L}_{\alpha\alpha^{\prime},\beta\beta^{\prime}}^{(0)}
\delta_{\alpha\beta}\delta_{\alpha^{\prime}\beta^{\prime}}
-J_{\alpha\alpha^{\prime},\beta\beta^{\prime}}\nonumber\\
&=&\left(i\omega_{\alpha\alpha^{\prime}}
+iL_{\alpha\alpha^{\prime}}\right)
\delta_{\alpha\beta}\delta_{\alpha^{\prime}\beta^{\prime}}
-J_{\alpha\alpha^{\prime},\beta\beta^{\prime}}\;.
\label{eq:qc-l}
\end{eqnarray}
Here, $\omega_{\alpha\alpha^{\prime}}=
\left(E_{\alpha}(R)-E_{\alpha^{\prime}}(R)\right)/\hbar
\equiv E_{\alpha\alpha^{\prime}}/\hbar$ and
\begin{equation}
iL_{\alpha\alpha^{\prime}}
=\frac{P}{M}\cdot\frac{\partial}{\partial R}
+\frac{1}{2}\left(F_{\alpha}+F_{\alpha^{\prime}}\right)
\frac{\partial}{\partial P}\;,
\label{eq:ilad}
\end{equation}
where
\begin{equation}
F_{\alpha}=
-\langle\alpha;R\vert\frac{\partial\hat{h}(R)}{\partial R}
\vert\alpha;R\rangle
\end{equation}
is the Hellmann-Feynman force for state $\alpha$.
The operator $J$ that describes nonadiabatic effects is
\begin{eqnarray}
J_{\alpha\alpha^{\prime},\beta\beta^{\prime}}
&=&-\frac{P}{M}\cdot d_{\alpha\beta}
\left(1+\frac{1}{2}S_{\alpha\beta}\cdot
\frac{\partial}{\partial P}\right)\delta_{\alpha^{\prime}\beta^{\prime}}
\nonumber\\
&&-\frac{P}{M}\cdot d_{\alpha^{\prime}\beta^{\prime}}^*
\left(1+\frac{1}{2}S_{\alpha^{\prime}\beta^{\prime}}^*\cdot
\frac{\partial}{\partial P}\right)\delta_{\alpha\beta}
\;,\nonumber\\
\label{eq:jad}
\end{eqnarray}
where $d_{\alpha\beta}=\langle\alpha;R\vert(\partial/\partial R)
\vert\beta;R\rangle$ is the nonadiabatic coupling vector
and
\begin{equation}
S_{\alpha\beta}=E_{\alpha\beta}d_{\alpha\beta}
\left(\frac{P}{M}\cdot d_{\alpha\beta}\right)^{-1}\;.
\end{equation}
Using Eqs.~(\ref{eq:ilad}) and ~(\ref{eq:jad}),
the equation of motion for the density matrix in the adiabatic basis
can be written explicitly as
\begin{eqnarray}
\partial_t\rho_{\alpha\alpha^{\prime}}
&=&
-i\omega_{\alpha\alpha^{\prime}}\rho_{\alpha\alpha^{\prime}}
-\frac{P}{M}\cdot\frac{\partial}{\partial R}\rho_{\alpha\alpha^{\prime}}
\nonumber\\
&& -\frac{1}{2}\left(F_{\alpha}+F_{\alpha^{\prime}}\right)\cdot
\frac{\partial}{\partial P}\rho_{\alpha\alpha^{\prime}}
\nonumber\\
&&-\sum_{\beta}\frac{P}{M}\cdot d_{\alpha\beta}
\left(1+\frac{1}{2}S_{\alpha\beta}\cdot\frac{\partial}{\partial P}\right)
\rho_{\beta\alpha^{\prime}}
\nonumber\\
&&-\sum_{\beta^{\prime}}\frac{P}{M}\cdot d_{\alpha^{\prime}\beta^{\prime}}^*
\left(1+\frac{1}{2}S_{\alpha^{\prime}\beta^{\prime}}^*
\cdot\frac{\partial}{\partial P}\right)
\rho_{\alpha\beta^{\prime}}
\;.\nonumber\\
\label{eq:rho-eq-ad}
\end{eqnarray}
The wave fields $\vert\psi^{\iota}(X)\rangle$ and $\langle\psi^{\iota}(X)\vert$
can be expanded in the adiabatic basis as
\begin{eqnarray}
\vert\psi^{\iota}(X)\rangle&=&\sum_{\alpha}\vert\alpha;R\rangle
\langle\alpha;R\vert\psi^{\iota}(X)\rangle
=\sum_{\alpha}C_{\alpha}^{\iota}\vert\alpha;R\rangle\nonumber\\
\langle\psi^{\iota}(X)\vert&=&\sum_{\alpha}\langle\psi^{\iota}\vert\alpha;R\rangle
\langle\alpha;R\vert
=\sum_{\alpha}\langle\alpha;R\vert C_{\alpha}^{\iota *}(X)
\;,\nonumber\\
\end{eqnarray}
and the density matrix in Eq.~(\ref{eq:rho-ansatz}) becomes
\begin{eqnarray}
\rho_{\alpha\alpha^{\prime}}(X,t)
&=&\sum_{\iota}w_{\iota}C_{\alpha}^{\iota}(X,t)C_{\alpha^{\prime}}^{\iota *}(X,t)
\;.
\label{eq:rho-ansatz-ad}
\end{eqnarray}
In order to find two separate equations for $C_{\alpha}^{\iota}$ and
$C_{\alpha^{\prime}}^{\iota *}$, one cannot insert Eq.~(\ref{eq:rho-ansatz-ad})
directly into Eq.~(\ref{eq:rho-eq-ad}) because of the presence
of the derivatives with respect to the phase space coordinates
$R$ ad $P$. One must set Eq.~(\ref{eq:rho-eq-ad}) into the
form of a multiplicative operator acting on $\rho_{\alpha\alpha^{\prime}}$.
To this end, for example, consider
\begin{eqnarray}
\frac{\partial}{\partial P}\rho_{\beta\alpha^{\prime}}
&=&\sum_{\gamma}\left(\frac{\partial}{\partial P}\rho_{\beta\gamma}
\right)\delta_{\gamma\alpha^{\prime}}
=\sum_{\gamma\mu}\left(\frac{\partial}{\partial P}\rho_{\beta\gamma}\right)
\rho_{\gamma\mu}^{-1}\rho_{\mu\alpha^{\prime}}
\nonumber\\
&=&\sum_{\mu}\frac{\partial(\ln\hat{\rho})_{\beta\mu}}{\partial P}
\rho_{\mu\alpha^{\prime}}\;.
\label{eq:der-mani}
\end{eqnarray}
Equation~(\ref{eq:der-mani}) shows how to transform formally
a derivative operator acting on $\hat{\rho}$ into a multiplicative operator
which, however, depends on $\hat{\rho}$ itself.
Therefore, Eq.~(\ref{eq:rho-eq-ad}) becomes
\begin{eqnarray}
\partial_t\rho_{\alpha\alpha^{\prime}}
&=&
-\frac{i}{\hbar}E_{\alpha}\rho_{\alpha\alpha^{\prime}}
+\frac{i}{\hbar}E_{\alpha^{\prime}}\rho_{\alpha\alpha^{\prime}}
\nonumber\\
&-&\sum_{\beta}\frac{P}{M}\cdot d_{\alpha\beta}\rho_{\beta\alpha^{\prime}}
-\sum_{\beta^{\prime}}\frac{P}{M}\cdot d_{\alpha^{\prime}\beta^{\prime}}^*
\rho_{\alpha\beta^{\prime}}
\nonumber\\
&-&\frac{1}{2}\sum_{\mu}\frac{P}{M}\cdot
\frac{\partial(\ln\rho)_{\alpha\mu}}{\partial R}
\rho_{\mu\alpha^{\prime}}
\nonumber\\
&-&\frac{1}{2}\sum_{\mu}\frac{P}{M}\cdot
\frac{\partial(\ln\rho)_{\mu\alpha^{\prime}} }{\partial R}
\rho_{\alpha\mu}
\nonumber\\
&-&\frac{1}{2}\sum_{\mu}F_{\alpha}
\frac{\partial(\ln\rho)_{\alpha\mu}}{\partial P}\rho_{\mu\alpha^{\prime}}
\nonumber\\
&-&\frac{1}{2}\sum_{\mu}F_{\alpha^{\prime}}\cdot
\frac{\partial(\ln\rho)_{\mu\alpha^{\prime}}}{\partial P}\rho_{\alpha\mu}
\nonumber\\
&-&\frac{1}{2}\sum_{\beta,\mu}\frac{P}{M}\cdot d_{\alpha\beta}
S_{\alpha\beta}\cdot\frac{\partial(\ln\hat{\rho})_{\beta\mu}}{\partial P}
\rho_{\mu\alpha^{\prime}}
\nonumber\\
&-&\frac{1}{2}\sum_{\beta^{\prime},\mu}\frac{P}{M}\cdot
d_{\alpha^{\prime}\beta^{\prime}}^* S_{\alpha^{\prime}\beta^{\prime}}^*
\cdot\frac{\partial(\ln\hat{\rho})_{\mu\beta^{\prime}} }{\partial P}
\rho_{\alpha\mu}
\;.\nonumber\\
\label{eq:rho-mani-ad}
\end{eqnarray}
Inserting the adiabatic expression for the density matrix,
given in Eq.~(\ref{eq:rho-ansatz-ad}), into Eq.~(\ref{eq:rho-mani-ad}),
one obtains, for each quantum state $\iota$, the following two coupled
equations
\begin{eqnarray}
\dot{C}_{\alpha}^{\iota}(X,t)
&=&
-\frac{i}{\hbar}E_{\alpha}C_{\alpha}^{\iota}(X,t)
-\sum_{\beta}\frac{P}{M}\cdot d_{\alpha\beta}C_{\beta}^{\iota}(X,t)
\nonumber\\
&-&\frac{1}{2}\sum_{\beta,\mu}\frac{P}{M}\cdot d_{\alpha\beta}
S_{\alpha\beta}\cdot\frac{\partial(\ln\hat{\rho})_{\beta\mu}}{\partial P}
C_{\mu}^{\iota}(X,t)
\nonumber\\
&-&\frac{1}{2}\sum_{\mu}\frac{P}{M}\cdot
\frac{\partial(\ln\rho)_{\alpha\mu}}{\partial R}C_{\mu}^{\iota}(X,t)
\nonumber\\
&-&\frac{1}{2}\sum_{\mu}F_{\alpha}
\frac{\partial(\ln\rho)_{\alpha\mu}}{\partial P}C_{\mu}^{\iota}(X,t)
\label{eq:c}\\
\dot{C}_{\alpha^{\prime}}^{\iota *}(X,t)
&=&
+\frac{i}{\hbar}E_{\alpha^{\prime}}C_{\alpha^{\prime}}^{\iota *}(X,t)
-\sum_{\beta^{\prime}}\frac{P}{M}\cdot d_{\alpha^{\prime}\beta^{\prime}}^*
C_{\beta^{\prime}}^{\iota *}(X,t)
\nonumber\\
&-&\frac{1}{2}\sum_{\beta^{\prime},\mu}\frac{P}{M}\cdot
d_{\alpha^{\prime}\beta^{\prime}}^* S_{\alpha^{\prime}\beta^{\prime}}^*
\cdot\frac{\partial(\ln\hat{\rho})_{\mu\beta^{\prime}} }{\partial P}
C_{\mu}^{\iota *}(X,t)
\nonumber\\
&-&
\frac{1}{2}\sum_{\mu}\frac{P}{M}\cdot
\frac{\partial(\ln\rho)_{\mu\alpha^{\prime}} }{\partial R}
C_{\mu}^{\iota *}(X,t)
\nonumber\\
&-&
\frac{1}{2}\sum_{\mu}F_{\alpha^{\prime}}\cdot
\frac{\partial(\ln\rho)_{\mu\alpha^{\prime}}}{\partial P}C_{\mu}^{\iota *}(X,t)
\;.\label{eq:cstar}
\end{eqnarray}
Quantum-classical averages of arbitrary observables can be calculated in the adiabatic as
\begin{equation}
\langle\hat{\chi}\rangle(t)
=
\sum_{\iota}w_{\iota}\sum_{\alpha\alpha^{\prime}}
\int dXC_{\alpha}^{\iota}(X,t)C_{\alpha^{\prime}}^{\iota *}(X,t)
\chi_{\alpha^{\prime}\alpha}(X)\;,
\label{eq:qc-ave-ad}
\end{equation}
where the coefficients $C_{\alpha}^{\iota}(X,t)$
and $C_{\alpha^{\prime}}^{\iota *}(X,t)$ are evolved according to Eqs.~(\ref{eq:c}) and~(\ref{eq:cstar}), respectively.
Equations~(\ref{eq:c}) and~(\ref{eq:cstar})
are non-linear equations which couple all the adiabatic states
used to analyze the system.

At this stage, a general discussion about such a non-linear character is
required. With a wide consensus, quantum mechanics is considered 
a linear theory.
This leads, for example, to the visualization
of  quantum transitions as instantaneous \emph{quantum jumps}.
The linearity of the theory also determines the need of considering
infinite perturbative series which must be re-summed in some way
in order to extract meaningful predictions.
Density Functional Theory is an example of a non-linear
theory~\cite{dft} but it is usually considered just as a computational tool.
However, there are other approaches to quantum theory
that represent interactions by an intrinsic non-linear scheme~\cite{mead}.
It is not difficult to see how this is possible.
Matter is represented by waves, these very same waves enter into the
definition of the fields defining their interaction~\cite{tomonaga}.
This point of view has been pursued by Jaynes~\cite{jaynes}
and Barut~\cite{barut}, among others.
These non-linear approaches depict quantum transitions
as abrupt but continuous events~\cite{mead} in which, to go from state
$\vert 1\rangle$ to state $\vert 2\rangle$, the system is
first brought by the interaction in a superposition
$\alpha\vert 1\rangle+\beta\vert 2\rangle$, and then,
as the interaction ends, it finally goes to state $\vert 2\rangle$.
It is understood that this is made possible by the non-linearity
of such theories because, instead, a linear theory would preserve
the superposition indefinitely.
Incidentally, the picture of the transition process just depicted
also emerges from the numerical implementation~\cite{kapral}
of the nonadiabatic quantum-classical
dynamics of phase space dependent operators~\cite{qc-bracket,kcmqc}:
The action of the operator $J$ in Eq.~(\ref{eq:qc-l}) can build
and destroy coherence in the system by creating
and destroying superposition of states.
As explained above, this is a feature of a non-linear theory.
Such a non-linear character is simply hidden in the operator
version of quantum-classical dynamics and clearly manifested
by the wave picture of the quantum-classical evolution,
which has been introduced in this paper.

Since Eqs.~(\ref{eq:c}) and~(\ref{eq:cstar}) are non-linear,
their numerical integration requires either to adopt an
iterative self-consistent procedure (according to which
one makes a first guess of $\rho_{\alpha\alpha^{\prime}}$,
as dictated by Eq.~(\ref{eq:rho-ansatz-ad}), calculates 
the evolved $C_{\alpha}^{\iota}(X,t)$
and $C_{\alpha^{\prime}}^{\iota *}(X,t)$,
and then goes into a recursive procedure
until numerical convergence is obtained)
or to choose a definite form for $\rho_{\alpha\alpha^{\prime}}^G$,
following physical intuition, and then calculating
the time evolution, according to the form of Eqs.~(\ref{eq:c}) and~(\ref{eq:cstar})
which is obtained by using $\rho_{\alpha\alpha^{\prime}}^G$.
This last method is already known within the 
Wigner formulation of quantum mechanics~\cite{lee}
as the method of \emph{Wigner trajectories}~\cite{wignertraj}.
It is also important to find some importance sampling scheme
for the phase space integral in Eq.~(\ref{eq:qc-ave-ad}).
Such sampling scheme may depend on the specific form $\chi_{\alpha\alpha^{\prime}}$ 
of the observable.
It is interesting to note that Eqs.~(\ref{eq:c}), (\ref{eq:cstar}),
and~(\ref{eq:qc-ave-ad}) can be used to address both equilibrium and
non-equilibrium problems on the same footing.
However, the dynamical picture provided by Eqs.~(\ref{eq:c}) and~(\ref{eq:cstar}) 
is very different both from that of the usual surface-hopping schemes~\cite{tully}
and from that of the nonadiabatic evolution of quantum-classical operators~\cite{kapral}.
In order to appreciate this,
for simplicity, one can consider a situation in which there is no thermal disorder
in the quantum degrees of freedom so that $\iota=1$: \emph{viz.},
the density matrix becomes that of a pure state
$\rho_{\alpha\alpha^{\prime}}(X,t)\to C_{\alpha}(X,t)C_{\alpha^{\prime}}^{*}(X,t)$.
Then, equations~(\ref{eq:c}) and~(\ref{eq:cstar}) remain unaltered
and one has just to remove the index $\iota$ from the coefficients.
Therefore, it can be realized that no classical trajectory propagation,
and no state switching
are involved by Eqs.~(\ref{eq:c}) and~(\ref{eq:cstar}).
Instead, in order to calculate averages 
according to Eq.~(\ref{eq:qc-ave-ad}),
one has to sample phase space points and integrate
the matrix equations.

In the next section, an equilibrium approximation of
Eqs.~(\ref{eq:c}) and~(\ref{eq:cstar}), along the lines 
followed by the method of \emph{Wigner trajectories}~\cite{wignertraj},
is given and applied, with good numerical results,
to the adiabatic and nonadiabatic dynamics of the spin-boson model.

\section{Wave dynamics of the spin-boson model}\label{sec:sb}

The theory developed in the previous sections can be applied
to simulate the relaxation dynamics of the spin-boson system~\cite{sb}.
This system has already been studied within the framework of quantum-classical
dynamics of operators in Ref.~\cite{qc-sb} and ``exact'' numerical results
were obtained at short-time by means of an iterative path integral
procedure developed by Nancy Makri and co-worker~\cite{makri}.
The short-time results of Ref.~\cite{sb} numerically coincide
with those obtained by the path integral calculation of Ref.~\cite{makri}.
However, as it is shown later by Fig.~\ref{fig:fig2},
the quantum-classical results of Ref.~\cite{sb} have some limitations concerning
the numerical stability of the algorithm beyong a certain time length.
Using the dimensionless variables of Ref.~\cite{qc-sb},
the quantum-classical Hamiltonian operator of the spin-boson system reads
\begin{eqnarray}
\hat{H}(X)&=&-\Omega\hat{\sigma}_x
+\sum_{j=1}^N\left(\frac{P_j^2}{2}+
\frac{1}{2}\omega_j^2R_j^2-c_j\hat{\sigma}_zR_j\right)
\nonumber\\
&=&
\hat{h}_s+H_b+\hat{V}_c(R)\;,
\end{eqnarray}
where $\hat{h}_s=-\Omega\hat{\sigma}_x$ is the subsystem Hamiltonian,
$H_b=\sum_{j=1}^NP_j^2/2+1/2\omega_j^2R_j^2=
\sum_{j=1}^NP_j^2/2+V_b(R)$ is the Hamiltonian of a 
classical bath of $N$ harmonic oscillators, and 
$\hat{V}_c(R)=-\sum_{j=1}^Nc_j\hat{\sigma}_zR_j=\gamma(R)\hat{\sigma}_z$
is the interaction between the subsystem and the bath.
An Ohmic spectral density is assumed for the bath.
Hence, denoting the Kondo parameter as $\xi_K$ and
the cut-off frequency as $\omega_{\rm max}$,
the frequencies of the oscillators are defined by
$\omega_j=-\ln(1-j\omega_0)$,
where $\omega_0=N^{-1}(1-\exp(-\omega_{\rm max}))$,
and the constants entering the coupling by
and $c_j=\sqrt{\xi_K\omega_0}~\omega_j$. 
The adiabatic eigenvalues and eigenvectors, respectively, are
\begin{equation}
E_{1,2}=V_b\mp\sqrt{\Omega^2+\gamma^2(R)}\;,
\end{equation}
\begin{eqnarray}
\vert 1;R\rangle&=& \frac{1}{\sqrt{2(1+G^2}}\left(\begin{array}{c} 1+G\\
1-G\end{array} \right)
\nonumber\\
\vert 2;R\rangle&=& \frac{1}{\sqrt{2(1+G^2}}\left(\begin{array}{c}-1+G\\
1+G\end{array} \right)\;,
\end{eqnarray}
where
\begin{equation}
G(R)=\gamma^{-1}(R)\left[-\Omega+\sqrt{\Omega^2+\gamma^2(R)}\right]\;.
\end{equation}
The coupling vector $d_{\alpha\alpha'}=\langle\alpha;R\vert\overrightarrow{\partial}
/\partial R\vert\alpha';R\rangle$ is
\begin{equation}
d_{12}=-d_{21}=(1+G^2)^{-1}\partial G/\partial R\;.
\end{equation}
Assuming an initially uncorrelated density matrix, where the bath is in 
thermal equilibrium and the subsystem is in state $\vert\uparrow\rangle$,
the initial quantum-classical density matrix in the adiabatic basis takes the form
\begin{equation}
\mbox{\boldmath$\rho$}(0)
=\mbox{\boldmath$\rho$}_s(0)\rho_b(X)\;,
\end{equation}
where
\begin{equation}
\mbox{\boldmath$\rho$}_s(0)=\frac{1}{2(1+G^2)}
\left(\begin{array}{cc}(1+G)^2 & 1-G^2 \\ 1-G^2 & (1-G)^2\end{array}\right)\;,
\end{equation}
and
\begin{eqnarray}
\rho_b(X)&=&\prod_{I=1}^N\frac{\tanh(\beta\omega_i/2)}{\omega_i}
\nonumber\\
&\times& \exp\left[-\frac{2\tanh(\beta\omega_i/2)}{\omega_i}
\left(\frac{P_i^2}{2}+\frac{\omega_i^2R_i^2}{2}\right)\right]\;.
\nonumber\\
\end{eqnarray}
The process of relaxation from the initial state can be followed by monitoring
the subsystem observables $\hat{\sigma_z}$, which in the adiabatic basis reads
\begin{equation}
\mbox{\boldmath$\sigma$}_z=\frac{1}{1+G^2}\left(
\begin{array}{cc} 2G & 1-G^2 \\ 1-G^2 & -2G\end{array}\right)\;.
\end{equation}

The adiabatic basis is real so that the initial density matrix
of the system can be written as
\begin{equation} 
\rho_{\alpha\alpha'}(X,0)=
\sum_{\alpha=1}^2 \psi_{\alpha}(X,0) \phi_{\alpha'}(X,0) \;,
\end{equation}
where
\begin{eqnarray}
\psi_{1}(X,0)= \phi_{1}(X,0)
&=&\sqrt{\rho_b(X)}\frac{1+G}{\sqrt{2(1+G^2)}}\;,
\nonumber\\
\\
\psi_{2}(X,0)= \phi_{2}(X,0)
&=&\sqrt{\rho_b(X)}\frac{1-G}{\sqrt{2(1+G^2)}}\;.
\nonumber\\
\end{eqnarray}
Such coefficients enter into the calculation of the observable
\begin{equation}
\langle\mbox{\boldmath$\sigma$}_z(t)\rangle
=
\sum_{\alpha\alpha'}\int dX \phi_{\alpha'}(X,t)\sigma_{z}^{\alpha'\alpha}(X)
\psi_{\alpha}(X,t)\;.
\label{eq:sigma-sb}
\end{equation}
The coefficients evolve in time according
to Eqs.~(\ref{eq:c}) and~(\ref{eq:cstar}),
where one must set $C_{\alpha}^{\iota}\equiv \psi_{\alpha}$
and $C_{\alpha'}^{\iota *}\equiv \phi_{\alpha'}$.
In order to devise an effective computational scheme
for  such equations, one could assume that the density matrix entering
Eqs.~(\ref{eq:c}) and~(\ref{eq:cstar}) is taken to be that at $t=\infty$,
when the total system (subsystem plus bath)
has reached thermal equilibrium.
The equilibrium quantum-classical density matrix is known
as a series expansion in $\hbar$~\cite{qc-stat}. If one makes the additional
assumption of complete decoherence at $t=\infty$,
only the ${\cal O}(\hbar^0)$ term can be taken
\begin{equation}
\rho_{e}^{(0)\alpha\alpha'}(X)
=Z_0^{-1}e^{-\beta(\sum_jP_j^2/2+E_{\alpha}(R))}\delta_{\alpha\alpha'}
\;,
\end{equation}
where $Z_0=\sum_{\alpha\alpha'}\int dX\rho_{e}^{(0)\alpha\alpha'}(X)$.
Then
\begin{eqnarray}
\frac{\partial\ln \rho_{e}^{(0)\alpha\alpha'} }{\partial R}
&=&-\beta\frac{\partial E_{\alpha}}{\partial R}\delta_{\alpha\alpha'}
\equiv \beta F_{\alpha}(R)\delta_{\alpha\alpha'}\;, 
\\
\frac{\partial\ln \rho_{e}^{(0)\alpha\alpha'} }{\partial P}
&=&-\beta P\delta_{\alpha\alpha'}\;.
\end{eqnarray}
Equations~(\ref{eq:c}) and~(\ref{eq:cstar}) become
\begin{eqnarray}
\frac{d}{dt}{\psi}_{\alpha}(X,t)
&=&
-iE_{\alpha}\psi_{\alpha}(X,t)\nonumber\\
&-&\sum_{\beta}P\cdot d_{\alpha\beta}
\left(1-\frac{\beta}{2}E_{\alpha\beta}\right)
\psi_{\beta}(X,t)
\label{eq:c2}\\
\frac{d}{dt}{\phi}_{\alpha^{\prime}}(X,t)
&=&
iE_{\alpha^{\prime}}\phi_{\alpha^{\prime}}(X,t)
\nonumber\\
&-&\sum_{\beta^{\prime}}P\cdot d_{\alpha^{\prime}\beta^{\prime}}
\left(1-\frac{\beta}{2}E_{\alpha^{\prime}\beta^{\prime}}\right)
\phi_{\beta^{\prime}}(X,t)
\;.
\nonumber\\
\label{eq:cstar2}
\end{eqnarray}
In Eqs.~(\ref{eq:cstar}) and~(\ref{eq:cstar2}) the terms
$\pm(\beta/2)P\cdot F_{\alpha} \psi_{\alpha}$ (and the analogous terms
with $\xi_{\alpha'}$) cancel each other.
In the adiabatic basis $d_{11}(R)=d_{22}(R)=0$. 
Hence, defining the matrix
\begin{eqnarray}
\mbox{\boldmath$\Sigma$}
&=&
\left[\begin{array}{cc} -iE_1 & -P\cdot d_{12}\left(1-\frac{\beta}{2}E_{12}\right)\\
P\cdot d_{12}\left(1+\frac{\beta}{2}E_{12}\right) & -iE_2 \end{array}\right]\;,
\nonumber\\
\end{eqnarray}
Equations~(\ref{eq:c2}) and~(\ref{eq:cstar2}) can be written as
\begin{eqnarray}
\frac{d}{dt}\left[\begin{array}{c}{\psi}_1 \\ {\psi}_2\end{array}\right]
&=&
\mbox{\boldmath$\Sigma$}\cdot
\left[\begin{array}{c}\psi_1 \\ \psi_2\end{array}\right]
\;, \quad
\frac{d}{dt}\left[\begin{array}{c}{\phi}_1 \\ {\phi}_2\end{array}\right]
=
\mbox{\boldmath$\Sigma$}^*\cdot
\left[\begin{array}{c}\phi_1 \\ \phi_2\end{array}\right]
\;,\nonumber\\
\label{eq:matrixSigma}
\end{eqnarray}
which can be integrated by means of the simple algorithm
$\mbox{\boldmath$\Psi$}(X,d\tau)=\mbox{\boldmath$\Psi$}(X,0)+
d\tau\mbox{\boldmath$\Theta$}(X,0)\cdot
\mbox{\boldmath$\eta$}(X,0)$, 
where $\mbox{\boldmath$\Psi$}=
(\mbox{\boldmath$\psi$},\mbox{\boldmath$\phi$})$
and 
$\mbox{\boldmath$\Theta$}=(\mbox{\boldmath$\Sigma$},
\mbox{\boldmath$\Sigma$}^*)$.
The phase space part of the initial values
of $\mbox{\boldmath$\psi$}$ and $\mbox{\boldmath$\phi$}$
can be used as the weight for sampling the coordinates $X$ 
entering the classical integral in Eq.~(\ref{eq:sigma-sb}).
Then, for each initial value $X$, Eqs.~(\ref{eq:matrixSigma})
must be integrated in time so that averages can be calculated.
It is worth to note that in such a wave scheme the Eulerian point
of view of quantum-classical dynamics~\cite{kapral,qc-sb}
is preserved. This is different from what happens
in the original operator approach~\cite{kapral,qc-sb},
where in order to devise an effective time integration scheme
by means of the Dyson expansion, one is forced to change
from the Eulerian point of view (according to which the
phase space point is fixed and the quantum degrees of freedom evolve in time
\emph{at} this fixed phase space point)
to the Lagrangian point of view, where phase space trajectories
are generated. Moreover, 
it must be noted that the numerical integration of Eqs.~(\ref{eq:matrixSigma})
provides directly the nonadiabatic dynamics without the need to introduce
surface-hopping approximations.

\begin{figure}
\resizebox{7cm}{4cm}{
\includegraphics* {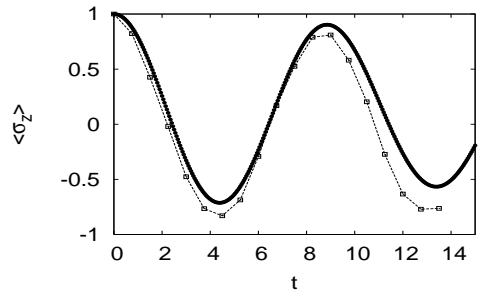}
}
\caption{Adiabatic dynamics of the spin-boson model.
$\beta=0.3$, $\Omega=1/3$, $\omega_{\rm max}=3$,
$\xi_K=0.007$, $N=200$. The black circles show the results of
the calculation with the theory proposed in this paper
while the light dashed line with squares shows, for comparison, 
the results of the calculation in Ref.~\cite{qc-sb}.
}
\label{fig:fig1}
\end{figure}

\begin{figure}
\resizebox{7cm}{4cm}{
\includegraphics* {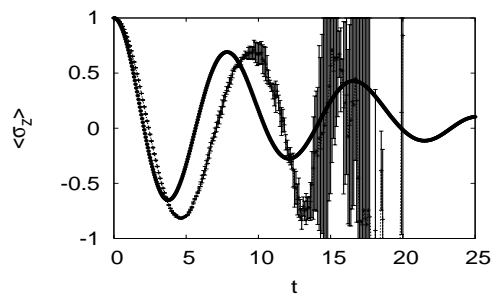}
}
\caption{Nonadiabatic dynamics of the spin-boson model.
$\beta=0.3$, $\Omega=1/3$, $\omega_{\rm max}=3$,
$\xi_K=0.007$, $N=200$. The black circles show the results of
the calculation with the theory proposed in this paper
while the line with error bars shows, for comparison, 
the results of the calculation in Ref.~\cite{qc-sb}.
These latter are indistinguishable, at short time,
from the ``exact'' results of Ref.~\cite{makri}, 
which were obtained by means of an iterative path integral
procedure.
}
\label{fig:fig2}
\end{figure}

In order to be able of comparing the results with those
presented in Ref.~\cite{qc-sb}, the numerical values
of the parameters specifying the spin-boson system
have been chosen to be $\beta=0.3$, $\Omega=1/3$, $\omega_{\rm max}=3$,
$\xi_K=0.007$, and $N=200$.
Figure~\ref{fig:fig1} shows the results 
in the adiabatic case, obtained by setting $d_{12}=0$
in Eqs.~(\ref{eq:matrixSigma}).
One can see that, in spite of the simple approximation
of the form of the density matrix made
in the equations of motion, the wave theory provides results
which are in good agreement with those obtained
with the operator approach of Ref.~\cite{qc-sb}.
Instead, Fig.~\ref{fig:fig2} shows the results
of the nonadiabatic calculation.
This is to be compared with the results of the operator theory~\cite{qc-sb}
(which are identical with the exact'' ones of Ref.~\cite{makri}).
Of course, since different ways of dealing with the nonadiabatic effects
are used in the two approaches the results do not need to be the same.
However, the results of the wave theory follow qualitatively
those of Ref.~\cite{qc-sb} while improving substantially
the statistical convergence and increasing the length of
the time interval spanned by a factor of $2-3$.
Such results are particularly encorauging and suggest
the possible application of the wave theory here proposed, for example,
to the calculation of nonadiabatic rate constants
of complex systems in the condensed phase~\cite{ksreview}.

\section{Conclusions}\label{sec:conclusions}

In this paper the approach to the quantum-classical mechanics of 
phase space dependent operators has been remodeled as a 
non-linear formalism for wave fields.
It has been shown that two coupled non-linear equations
for phase space dependent wave fields correspond 
to the single equation for the quantum-classical density matrix
in the operator scheme of motion.
The equations of motion for the wave fields have been
re-expressed by means of a suitable bracket
and it has been shown that the emerging formalism
generalizes within a non-Hamiltonian framework 
the non-linear quantum mechanical formalism that has been
proposed recently by Weinberg.
Finally, the non-linear wave equations have been represented
into the adiabatic basis and have been applied, after
a suitable equilibrium approximation, to the numerical study
of the adiabatic and nonadiabatic dynamics of the spin-boson model.
Good results have been obtained.
In particular, the time interval that can be spanned by 
the nonadiabatic calculation within the wave scheme of motion
turns out to be a factor of two-three longer than that
accessible within the operator scheme of motion.
This encourages one to pursue the application
of the wave scheme of motion to the calculation
of correlation functions for systems in the condensed phase.
Future works will be specifically devoted  to such an issue.

\vspace{1cm}

\noindent
{\bf Acknowledgment}

\noindent
I acknowledge Professor Kapral for suggesting
the possibility of mapping the quantum-classical 
dynamics of operators into a wave scheme of motion. 
I am also very grateful to Professor P. V. Giaquinta
for continuous encouragement and suggestions.
Finally, discussions with Dr Giuseppe Pellicane
during the final stage of this work are gratefully
acknowledged.

\appendix

\section{Weinberg's formalism}\label{app:weinberg}

Consider a quantum system in a state described by 
the wave fields $|\Psi\rangle$ and $\langle\Psi|$, where Dirac's bra-ket
notation is used to denote $\Psi(r)\equiv \langle r|\Psi\rangle$
and $\Psi^*(r)\equiv \langle \Psi| r\rangle$.
Observables are defined by functions of the type
\begin{equation}
a=\langle\Psi|\hat{A}|\Psi\rangle\;,
\end{equation}
where the operators are Hermitian, $\hat{A}=\hat{A}^{\dag}$.
Weinberg's formalism can be introduced by defining Poisson brackets in terms
of the wave fields $|\Psi\rangle$ and $\langle\Psi|$.
To this end, one considers the wave fields as ``phase space'' coordinates
$\mbox{\boldmath$\zeta$}\equiv(|\Psi\rangle , \langle\Psi|)$, so that $\zeta_1=|\Psi\rangle$
and $\zeta_2=\langle\Psi|$,
and then introduce brackets of observables as
\begin{eqnarray}
\{a,b\}_{\mbox{\tiny\boldmath$\cal B$}}&=&\sum_{\alpha=1}^2\frac{\partial a}{\partial \zeta_{\alpha}}
{\mathcal B}_{\alpha\beta}\frac{\partial b}{\partial \zeta_{\beta}}\;.
\label{eq:poissonbracket}
\end{eqnarray}
The bracket in Eq.~(\ref{eq:poissonbracket}) defines a Lie algebra
and a Hamiltonian systems. Typically, the Jacobi relation
is satisfied, \emph{i.e.}
$
{\cal J}=\left\{a,\left\{b,c\right\}_{\mbox{\tiny\boldmath$\cal B$}}
\right\}_{\mbox{\tiny\boldmath$\cal B$}}
+\left\{c\left\{a,b\right\}_{\mbox{\tiny\boldmath$\cal B$}}
\right\}_{\mbox{\tiny\boldmath$\cal B$}}
+\left\{b,\left\{c,a\right\}_{\mbox{\tiny\boldmath$\cal B$}}
\right\}_{\mbox{\tiny\boldmath$\cal B$}}=0$.
In order to obtain the usual quantum formalism,
one can introduce the Hamiltonian functional
in the form
\begin{equation}
{\cal H}[|\psi\rangle , \langle\psi|]\equiv {\cal H}[\mbox{\boldmath$\zeta$}]
= \langle\psi|\hat{H}|\psi\rangle \;, 
\label{eq:h_qm}
\end{equation}
where $\hat{H}$ is the Hamiltonian operator of the system. 
Equations of motion for the wave fields
can be written in compact form as 
\begin{equation}
\frac{\partial\mbox{\boldmath$\zeta$}}{\partial t}=\frac{i}{\hbar}
\{ {\cal H}[\mbox{\boldmath$\zeta$}] , \mbox{\boldmath$\zeta$} \}_{\mbox{\tiny\boldmath$\cal B$}} \;.
\label{eq:wein_eqofm}
\end{equation}
The compact form of Eq.~(\ref{eq:wein_eqofm})
can be set into an explicit form as
\begin{eqnarray}
\frac{\partial}{\partial t}|\Psi\rangle&=&\frac{i}{\hbar}
\frac{\partial{\cal H}}{\partial\langle\Psi|}
{\mathcal B}_{21}
\label{eq:wein_eqofm1}
\\
\frac{\partial}{\partial t} \langle\Psi|
&=&\frac{i}{\hbar}
\frac{\partial{\cal H}}{\partial\vert\Psi\rangle}
{\mathcal B}_{12}
\label{eq:wein_eqofm2}
\;.
\end{eqnarray}     
It is easy to see that, when the Hamiltonian function
is chosen as in Eq.~(\ref{eq:h_qm}),
Eq.~(\ref{eq:wein_eqofm}), or its explicit 
form~(\ref{eq:wein_eqofm1}-\ref{eq:wein_eqofm2}), gives
the usual formalism of quantum mechanics.
It is worth to remark that
in order not to alter gauge invariance,  the Hamiltonian
and the other observables must obey the homogeneity condition:
\begin{equation}
{\cal H}=\langle\Psi|(\partial{\cal H}/\partial\zeta_2)\rangle
=\langle(\partial{\cal H}/\partial\zeta_1)|\Psi\rangle
\;.\label{eq:homogeneity}
\end{equation}
Weinberg showed how the formalism above sketched can be generalized in order
to describe non-linear effects in quantum mechanics~\cite{weinberg}.
To this end, one must
maintain the homogeneity condition, Eq.~(\ref{eq:homogeneity}),
on the Hamiltonian but relax the constraint which assumes that
the Hamiltonian must be a bilinear function of the wave fields.
Thus, the Hamiltonian can be a general function given by
\begin{equation}
\tilde{\cal H}=\sum_{i=1}^n\rho^{-i}{\cal H}_i\;,
\end{equation}
where $n$ is arbitrary integer that fixes the order of the correction,
${\cal H}_0=h$, and
\begin{eqnarray}
{\cal H}_1&=&\rho^{-1}\int dr dr'dr''dr'''\Psi^*(r)\Psi^*(r')
\nonumber\\
&\times&
G(r,r',r'',r''')\Psi(r'')\Psi(r''')\;,
\end{eqnarray}
with analogous expressions for higher order terms.
Applications and thorough discussions of the above formalism
can be found in Ref.~\cite{weinberg}.

Once Weinberg's formalism is expressed by means of
the symplectic form in Eq.~(\ref{eq:wein_eqofm}),
it can be generalized very easily in order to obtain a
non-Hamiltonian quantum algebra.
To this end, one can substitute the antisymmetric
matrix $\mbox{\boldmath$\cal B$}$ with another antisymmetric matrix
$\mbox{\boldmath$\Omega$}=\mbox{\boldmath$\Omega$}[\mbox{\boldmath$\zeta$}]$, whose
elements might be functionals of 
$\mbox{\boldmath$\zeta$}\equiv(\vert\Psi\rangle,\langle\Psi\vert)$
obeying the homogeneity condition in Eq.~(\ref{eq:homogeneity}).
By means of $\mbox{\boldmath$\Omega$}$ a non-Hamiltonian bracket
$\left\{ \ldots ,\ldots \right\}_{\mbox{\tiny\boldmath$\Omega$}}$
can be defined as 
\begin{eqnarray}
\left\{ a,b\right\}_{\mbox{\tiny\boldmath$\Omega$}}&=&
\sum_{\alpha=1}^2\frac{\partial a}{\partial\zeta{\alpha}}
{\Omega}_{\alpha\beta}[\zeta]\frac{\partial b}{\partial\zeta{\beta}}\;.
\label{eq:nhbracket}
\end{eqnarray}
In general, the bracket in Eq.~(\ref{eq:nhbracket}) does no longer satisfy
the Jacobi relation
\begin{equation}
{\cal J}=
\left\{ a,\left\{ b,c\right\}_{\mbox{\tiny\boldmath$\Omega$}}
\right\}_{\mbox{\tiny\boldmath$\Omega$}} 
+\left\{ c\left\{ a,b\right\}_{\mbox{\tiny\boldmath$\Omega$}}
\right\}_{\mbox{\tiny\boldmath$\Omega$}}
+\left\{ b,\left\{ c,a\right\}_{\mbox{\tiny\boldmath$\Omega$}}
\right\}_{\mbox{\tiny\boldmath$\Omega$}} 
\neq 0\;.\label{eq:njacobi}
\end{equation}
Thus, non-Hamiltonian equations of motion 
can be written as
\begin{equation}
\frac{\partial\mbox{\boldmath$\zeta$}}{\partial t}=\frac{i}{\hbar}
\left\{ {\cal H},\mbox{\boldmath$\zeta$}\right\}_{\mbox{\tiny\boldmath$\Omega$}}
\;.
\label{eq:wein_nheqofm}
\end{equation}
In principle, the non-Hamiltonian theory, specified by 
Eqs.~(\ref{eq:nhbracket}),~(\ref{eq:njacobi}), 
and~(\ref{eq:wein_nheqofm}),
can be used
to address the problem of non-linear correction to quantum mechanics,
as it was done in Refs.~\cite{weinberg}.
In the present paper, it has been shown that
such a non-Hamiltonian and non-linear
version of quantum mechanics is already implied when
one formulates quantum-classical dynamics
of operators by means of suitable brackets.
As a matter of fact, it was shown that the quantum-classical theories
of Refs.~\cite{qc-bracket,kcmqc} can be mapped onto
a wave formalism which has precisely the same
form specified by Eqs.~(\ref{eq:nhbracket}),~(\ref{eq:njacobi}),
and~(\ref{eq:wein_nheqofm}).


\end{document}